\documentclass[twocolumn, superscriptaddress, secnumarabic, amssymb, nobibnotes, nofootinbib, aps, prd]{revtex4}
\usepackage{amsmath}
\usepackage{amsfonts}
\usepackage{amssymb}
\usepackage{comment}
\usepackage{float}
\usepackage{caption}
\usepackage{subcaption}
\usepackage{graphicx}
\usepackage{appendix}
\usepackage{color}
\usepackage{tikz}
\usetikzlibrary{quotes, angles}
\usepackage[unicode=true,
 bookmarks=false,
 breaklinks=false,pdfborder={0 0 1},backref=false,colorlinks=true]
 {hyperref}
\hypersetup{
 urlcolor=green,linkcolor=blue,citecolor=red}
 
\usepackage{soul}
\setstcolor{blue}

\begin{document}

\title{Quantum Gravitational Sensor for Space Debris}

\author{Meng-Zhi Wu}
\affiliation{Van Swinderen Institute, University of Groningen, 9747 AG, The Netherlands}

\author{Marko Toro\v{s}}
\affiliation{School of Physics and Astronomy, University of Glasgow, Glasgow, G12 8QQ, UK}

\author{Sougato Bose}
\affiliation{Department of Physics and Astronomy, University College London, Gower Street, WC1E 6BT London, United Kingdom}

\author{Anupam Mazumdar}
\affiliation{Van Swinderen Institute, University of Groningen, 9747 AG, The Netherlands}


\begin{abstract}
Matter-wave interferometers have fundamental applications for gravity experiments such as testing the equivalence principle and the quantum nature of gravity. In addition, matter-wave interferometers can be used as quantum sensors to measure the local gravitational acceleration caused by external massive moving objects, thus lending itself for technological applications. In this paper, we will establish a three dimensional model to describe the gravity gradient signal from an external moving object, and theoretically investigate the achievable sensitivities using the matter-wave interferometer based on the Stern-Gerlach set-up. As an application we will consider the Mesoscopic Interference for Metric and Curvature (MIMAC) and Gravitational wave detection scheme [New J. Phys. 22, 083012 (2020)] and quantify its sensitivity to gravity gradients using frequency-space analysis. We will consider objects near Earth-based experiments and space debris in proximity of satellites and estimate the minimum detectable mass of the object as a function of their distance, velocity, and orientation. 
\end{abstract}

\maketitle

\section{Introduction}\label{section1}

Interferometry has many salient applications~\cite{Bongs_2004} in gravity experiments such as testing the equivalence principle~\cite{PhysRevLett.120.183604, PhysRevLett.125.191101,Bose:2022czr} and measuring the Earth's gravitational acceleration~\cite{Peters1999,Marshman:2018upe,Chiao:2003sa,Roura:2004se, Foffa:2004up,Dimopoulos:2008sv,Dimopoulos:2007cj,Tino:2019tkb, Asenbaum:2020era, Graham:2012sy}. The seminal works on neutron interferometry~\cite{PhysRevLett.34.1472,werner1979effect,9780198712510.001.0001} motivated a series of matter-wave interferometers~\cite{Nesvizhevsky2002, science.1135459, asenbaum2017phase,overstreet2022observation} as well as led to more recent developments in photon interferometry~\cite{Bertocchi_2006, Fink2017, PhysRevLett.123.110401, torovs2020revealing,cromb2022controlling,torovs2022generation}.

One of the latest quests is to build a matter-wave interferometer with nanoparticles to test the quantum nature of gravity in a laboratory~\cite{PhysRevLett.119.240401,ICTS} (for a related work see~\cite{PhysRevLett.119.240402}). The scheme relies on two masses, each prepared in a spatial superposition, and placed at distances where they couple gravitationally, but still sufficiently far apart that all other interactions remain suppressed. If gravity is a bonafide quantum entity, and not a classical real-valued field, then the two masses will entangle~\cite{Marshman:2019sne,Bose:2022uxe,Christodoulou:2022vte,Danielson:2021egj}. To test the quantum nature of gravity we will need particles of mass $\sim10^{-14}-10^{-15}$\,kg, an interferometric scheme for preparing large superposition sizes $\sim100\,\rm\mu m$, and exquisite experimental control to guarantee coherence times of $\sim1$\,s~\cite{PhysRevLett.119.240401, Pino_2018, PhysRevA.102.062807, Toros:2020dbf, Tilly:2021qef,Schut:2021svd, Nguyen2020, PhysRevA.102.022428}. 

One of the most promising approaches towards interferometry with nanoparticles is based on the Stern-Gerlach (SG) apparatus~\cite{friedrich2021molecular}. SG interferometers have been already experimentally realized using an atom chip \cite{Keil2021}, with the half-loop~\cite{Machluf2013} and full-loop~\cite{SGI_experiment} configurations achieving the superposition size of 3.93\,$\rm\mu m$ and $0.38\,\rm\mu m$ in the experimental time of 21.45\,ms and 7\,ms,  respectively~\cite{SGI_experiment}. This basic SG scheme can be adapted to the mass range of nanoparticles using nanodiamond like materials with embedded nitrogen vacancy (NV) centers. Such a system has an internal spin degree of freedom and can thus be placed in a large spatial superposition using the SG setup~\cite{PhysRevLett.119.240401, Marshman:2021wyk, Zhou:2022frl, Zhou:2022jug,Zhou:2022epb}.

One of the main challenges of nanocrystal matter-wave interferometry is to tame the numerous decoherence and noise sources. Common sources for the loss of visibility, such as the ones arising from residual gas collisions and environmental photons, can be attenuated by vacuum and low-temperature technologies~\cite{Pino_2018, PhysRevA.102.062807, Toros:2020dbf, Tilly:2021qef,Schut:2021svd, Nguyen2020, PhysRevA.102.022428}. In addition, the spin decoherence should also been taken into account, i.e., the Humpty-Dumpty effect~\cite{Englert1988, Schwinger1988, PMID:9902333,Zhou:2022frl,Japha:2022xyg}, with methods to extend the spin coherence time, as well as tackle the Majorana spin-flip, under development~\cite{Inguscio:2007xi, Marshman:2021wyk,Zhou:2022frl}. Moreover, there are also a series of gravitational channels for decoherence; the emission of gravitons is negligible~\cite{torovs2020loss}, decoherence induced by the gravitational interaction with the experimental apparatus can be reduced using a hierarchy of distances~\cite{Gunnink:2022ner}, and gravity gradient noise (GGN) can be mitigated with an exclusion zone~\cite{Toros:2020dbf}. GGN is equally important for the gravitational wave observatories~\cite{Harms2019, PhysRevD.103.103017} such as LIGO~\cite{PhysRevD.58.122002, PhysRevD.60.082001, PhysRevD.98.083019}, Virgo~\cite{Beccaria:1998ap, Acernese_2014}, KAGRA~\cite{KAGRA}, LISA~\cite{Lisa406065, PhysRevD.70.063512, PhysRevD.106.063015, Adams:2004pk} and Einstein Telescope\cite{Bader_2022}, in particular at the low frequencies.

In this work, we will investigate the possibility of using the nanoparticle matter-wave interferometer as a \emph{gravity gradient quantum sensor}. We will estimate the required sensitivities to detect the motion of external objects flying at small and large impact parameters and with varying velocities. Such a device can be regarded as a quantum sensor, such as accelerometers, gravimeters and gradiometers~\cite{doi:10.1063/1.1150092, Marshman_2020, Qvarfort2018, PhysRevA.96.043824, rademacher2020quantum}.
 
We will first make a brief review about sensing with matter-wave interferometers in the language of Feynman's path integral approach (Sec.~\ref{section2}). As will be shown, the phase fluctuation density in the frequency space can be factorized into a noise part (described by the corresponding power spectrum density) multiplied by the trajectory part (described by the so-called transfer function). Then, we will establish a three dimensional model for the GGN as a signal caused by moving the external objects, in particular, obtaining the relation between the local acceleration noise and phase fluctuation (Sec.~\ref{3D}). We will also show that it recovers the two-dimensional model of Ref.~\cite{PhysRevD.30.732} in a specific limit (see Appendix~\ref{appendix}). We will apply our model to evaluate the possibility of tracking slow moving matter in an earth-based laboratories and space debris in the proximity of satellites using the Mesoscopic Interference for Metric and Curvature (MIMAC) and Gravitational wave interferometer \cite{Marshman:2018upe} (Sec.~\ref{nearGGN}), and give a comparison to the quantum gravity induced-entanglement of masses (QGEM) which involves dual interferometer \cite{PhysRevLett.119.240401, ICTS,Toros:2020dbf} (see Appendix~\ref{AppendixB}).

\section{Noises in the Matter-wave Interferometry}\label{section2}

In this section, we will give a brief pedagogical introduction to the matter-wave sensing with a nanoparticles. According to Feynman's path integral method, the quantum phase along each path can be obtained from the action, and the signal in the experiment is described by the phase difference~\cite{storey1994feynman}:
\begin{equation}
\begin{split}
    \phi_0 &= \phi_R-\phi_L \\ 
        &= \frac{1}{\hbar}\int_{t_i}^{t_f}L_{\rm R}\left[x_{\rm R},\dot x_{\rm R}\right]-L_{\rm L}\left[x_{\rm L},\dot x_{\rm L}\right]dt,
\end{split}
\end{equation}
where $t_i$ and $t_f$ are the time of splitting and recombination of the two beams, $L_{\rm L,R}$ is the Lagrangian of the left and right arm which is a functional of the coordinate $x_{\rm L,R}\equiv x_{\rm L,R}(t)$ and the velocity $\dot x_{\rm L,R}\equiv x_{\rm L,R}(t)$. Supposing that the Lagrangian can be expanded as a Taylor series in $x_{\rm L,R}$, and that the noises can be described as the fluctuation of the coefficients, we find:
\begin{equation}
\begin{split}
    L_{L,R}\left[x_{\rm L,R},\dot x_{\rm L,R}\right] &= \frac{1}{2}m_0\dot x_{\rm L,R}^2\\ &-m_0a_{0; \rm L,R}x_{\rm L,R}-\frac{1}{2}m_0\omega_{0; \rm L,R}^2x_{\rm L,R}^2\\ &-m_0a_{\rm noise}x_{\rm L,R}-\frac{1}{2}m_0\omega_{\rm noise}^2x_{\rm L,R}^2\\ &+\mathcal{O}(x_{\rm L,R}^3),
\end{split}
\end{equation}
where $m_0$ is the mass of the interferometer, $a_{0; \rm L,R}$ and $\omega_{0; \rm L,R}^2$ are controlled by the experiment, and $a_{\rm noise}\equiv a_{\rm noise}(t)$ and $\omega_{\rm noise}^2\equiv \omega_{\rm noise}^2(t)$ are time-varying stochastic quantities. In particular, the GGN will be described by the quadratic term, so we will focus on $\omega_{\rm noise}^2$ in the rest of this section. In principle, $a_{\rm noise}$ and noises coupling higher order terms $\mathcal{O}(x_{\rm L,R}^3)$  can be studied in the same way.  Since the noise can be modelled as a fluctuation in the Lagrangian, it will contribute to a fluctuation in the phase difference $\phi_0 = \phi_R-\phi_L$, given by
\begin{equation}\label{delta-phi_0}
    \delta\phi_0 = \frac{m_0}{2\hbar}\int_{t_i}^{t_f}\omega_{\rm noise}^2(x_R^2-x_L^2)dt.
\end{equation}

Experimentally measurable statistical quantities are obtained by taking the average value  $\mathbb{E}[\;\cdot\;]$
\footnote{The symbol $\mathbb{E}[\;\cdot\;]$ represents the statistical average of a stochastic quantity, i.e., the average over different realizations of the noise. However, for a time-varying ergodic noise, the averaging can be also performed in time using a single realization of the noise. For example, the average of a time-varying stochastic quantity $\delta\phi(t)$ can be formulated as $$\mathbb{E}[\delta\phi]=\frac{1}{T}\int_0^T\delta\phi(t)dt,$$ where $T$ should be much longer than any time scale characterizing the statistical properties of the noise. More pedagogic materials can be found in \cite{ingle2005statisical}.}. The mean value of the noise $\mathbb{E}[\omega_{\rm noise}^2(t)]$ can be assumed to be zero by adding an offset on the baseline of the signal in experiments
\footnote{The baseline (i.e., the zero-point) of the phase has to be calibrated before the experiment starts, so the contribution of the mean value of every noise will be taken into account in the offset of the baseline. Therefore, the mean value of a noise $\mathbb{E}[\omega_{\rm noise}^2(t)]$ can be always assumed to be zero.}.
The autocorrelation function $\mathbb{E}[\omega_{\rm noise}^2(t_1)\omega_{\rm noise}^2(t_2)]$ can be related to the Fourier transformation of the corresponding power spectrum density (PSD) of the noise, denoted as $S_{\rm noise}(\omega, t)$, using the Wiener-Khinchin theorem. We further suppose the noise is stationary (i.e., its properties do not change over time), such that the PSD becomes time-independent $S_{\rm noise}(\omega, t)=S_{\rm noise}(\omega)$ (see for example ~\cite{chatfield10analysis}).

Summarizing, the noise $\omega_{\rm noise}^2(t)$ is characterised by the following statistical quantities:
\begin{equation}\label{Noise condition}
\begin{aligned}
    \mathbb{E}[\omega_{\rm noise}^2(t)] &= 0, \\
    \mathbb{E}[\omega_{\rm noise}^2(t_1)\omega_{\rm noise}^2(t_2)] &= \frac{1}{2\pi}\int_{\omega_{\rm min}}^\infty S_{\rm noise}(\omega)e^{i\omega(t_1-t_2)}d\omega.
\end{aligned}
\end{equation}
Here, we have  introduced a lower bound on the integral as $\omega_{\rm min}$ as a cut-off to avoid possible divergence in the integral. This lower bound can be assumed to be determined by the total experiment time $t_{\rm exp}=t_f-t_i$, i.e. $\omega_{\rm min}=2\pi/t_{\rm exp}$, which physically means that the interferometer is not sensitive to the frequencies with a period longer than the total experimental time. This infrared dependency on the cut-off relies also on a specific PSD. For our purpose, as we shall see we can take $\omega_{\rm min}\approx 0$.

By using Eqs.~\eqref{delta-phi_0} and \eqref{Noise condition}, we can find the average value of the phase fluctuation vanishes, while the variance is given by
\begin{equation}\label{phase-fluctuation}
    \Gamma_{\rm noise} \equiv \mathbb{E}[(\delta\phi_0)^2] = \frac{1}{2\pi}\left(\frac{m_0}{2\hbar}\right)^2 \int_{\omega_{\rm min}}^\infty S_{\rm noise}(\omega)F(\omega)d\omega,
\end{equation}
where $F(\omega)$ is defined by
\begin{alignat}{2}\label{shape-function}
    F(\omega) = && \int dt_1 \int dt_2 \left(x_R^2(t_2)-x_L^2(t_2)\right) \nonumber\\
     && \left(x_R^2(t_1)-x_L^2(t_1)\right) e^{i\omega(t_1-t_2)}.
\end{alignat}
Since $F(\omega)$ only depends on the trajectories of the two arms, we will call it the \emph{transfer function} of the interferometer\cite{Greve2022}, which means it transfers the PSD of the noise into the phase fluctuation of the interferometer. Mathematically, the double integral in $t_1$ and $t_2$ in Eq.\,(\ref{shape-function}) can be transformed to a product of two single integrals, so the transfer function $F(\omega)$ can be simplified as
\begin{equation}\label{F-equivalent-expression}
    F(\omega) = \left|\int \mathrm{e}^{\mathrm{i}\omega t}(x_R^2(t)-x_L^2(t))dt\right|^2 
\end{equation}
According to expression Eq.\,(\ref{F-equivalent-expression}), the transfer function $F(\omega)$ is the modulus square of a complex number integration, so it is always a real valued function. 

In the low-frequency regime, $\omega\ll2\pi/t_{\rm exp}$ (although this region is negligible according to the lower cut-off of the Fourier transformation), the factor $e^{i\omega t}$ in the first expression approximately equals one, then $F(\omega)$ approximately equals $\left(\int(x_R^2(t)-x_L^2(t))dt\right)^2$, which is independent of the frequency $\omega$. 

For the high-frequency noise, we can write the integrand $x_R^2(t)-x_L^2(t)$ into a polynomial series of $t$, i.e., $x_R^2(t)-x_L^2(t)=\sum_{n}^\infty c_nt^n$ of which each term will contribute a factor $\omega^{-n}$  after the integration in Eq.~(\ref{F-equivalent-expression}). So, $F(\omega)$ decreases in the high-frequency region as $\omega^{-k}$, where $k$ depends on the leading order $n$ of the polynomial expansion of $x_R^2(t)-x_L^2(t)$.

Therefore, the total phase fluctuation, $\Gamma_{\rm noise}$, is dominated by the lower frequency region, and sensitive to the lower bound $\omega_{\rm min}=2\pi/t_{\rm exp}$ of the integration, see Eq.(\ref{phase-fluctuation}). In particular, the shorter experimental time $t_{\rm exp}$ is, the larger the integral bound $\omega_{\rm min}$ is, and hence the smaller will be the total phase fluctuation, $\Gamma_{\rm noise}$. 
\begin{figure}
    \begin{tikzpicture}
    \draw[->, gray] (-2.5,0)--(3,0) node[right, black] {$x$};
    \draw[->, gray] (0.75,-0.5)--(0.75,4.5) node[above, black] {$t$};

    \draw (1.5,-0.25)--(1.5,0.25);
    \draw[thick, densely dotted] (1.5,0.25) .. controls (1.5,1) and (2,0.5) .. (2.5,1.5);
    \draw[thick, densely dotted] (2.5,1.5)--(2.5,2.25);
    \draw[thick, densely dotted] (2.5,2.25) .. controls (2,3.25) and (1.5,2.75) .. (1.5,3.5);
    \draw[thick, densely dotted] (1.5,0.25)--(1.5,3.5);
    \draw (1.5,3.5)--(1.5,4);
    
    \draw[fill=gray] (1.5,-0.25) circle (0.04);
    \draw[fill=gray] (1.5,0.25) circle (0.04);
    \draw[fill=gray] (1.5,1.7) circle (0.04);
    \draw[fill=gray] (2.5,1.7) circle (0.04);
    \draw[fill=gray] (1.5,3.5) circle (0.04);
    \draw[fill=gray] (1.5,4) circle (0.04);

    \draw[<->] (1.55,1.7)-- node[above]{$\Delta x$} (2.45,1.7);
    \draw[dashed] (-2.1,0.25)--(2.75,0.25);
    \draw[dashed] (-2.1,1.5)--(2.75,1.5);
    \draw[dashed] (-2.1,2.25)--(2.75,2.25);
    \draw[dashed] (-2.1,3.5)--(2.75,3.5);
    \draw[<->] (-1.9,0.3)-- node[left]{$2t_a$} (-1.9,1.45);
    \draw[<->] (-1.9,1.55)-- node[left]{$t_e$} (-1.9,2.2);
    \draw[<->] (-1.9,2.3)-- node[left]{$2t_a$} (-1.9,3.45);

    \node[right] at (-1.65, 0.9) {creation}; 
    \node[right] at (-1.75, 1.875) {free flight}; 
    \node[right] at (-1.8, 2.9) {recombination}; 
    
\end{tikzpicture}
    \caption{\small {The figures is the illustration of the paths of the two arms of the interferometer. The acceleration direction of the right arm is along "+" direction of the x-axis during the time range $[0,t_a]$ and $[3t_a+t_e,4t_a+t_e]$, while it is along "-" direction during $[t_a,2t_a]$ and $[2t_a+t_e,3t_a+t_e]$. In the interval $[2t_a,2t_a+t_e]$ the right paths follow geodesic motion, while the motion of the left arm is purely geodesic.  A single interferometer must be asymmetric to be sensitive to the GGN as one can always choose the origin of the harmonic trap generated by the GGN to be at the center of the two paths (a single symmetric interferometer would thus acquire only a global phase from any harmonic perturbation as the two paths would acquire exactly the same phase). We also assume that the setup is freely falling under gravity.}}
    \label{SGI}
\end{figure}
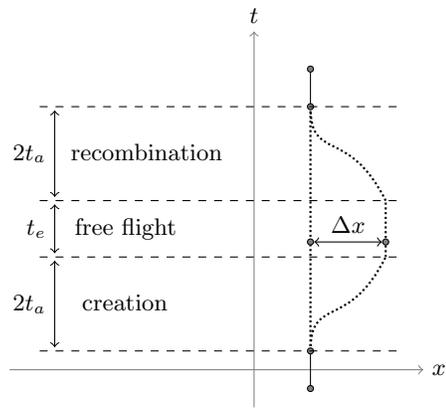
\begin{figure*}
  	\begin{subfigure}[]{0.3\textwidth}
        \includegraphics[width=\textwidth]{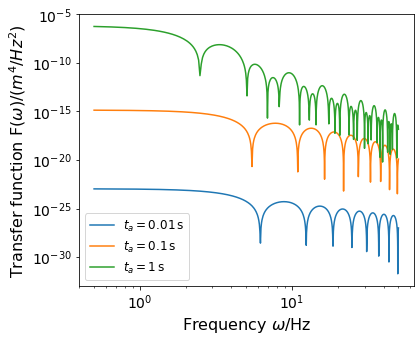}
        \subcaption*{(a)}
        \label{Fgg-symmetric-ta}
    \end{subfigure}
    \begin{subfigure}[]{0.3\textwidth}
        \includegraphics[width=\textwidth]{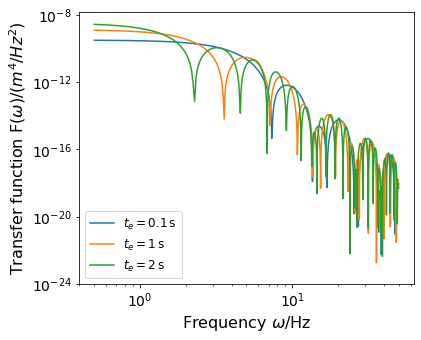}
        \subcaption*{(b)}
        \label{Fgg-symmetric-te}
    \end{subfigure}
    \begin{subfigure}[]{0.3\textwidth}
        \includegraphics[width=\textwidth]{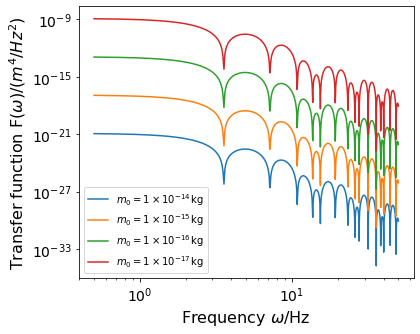}
        \subcaption*{(c)}
        \label{Fgg-symmetric-mass}
    \end{subfigure}
    \caption{
			\small {We have shown the transfer function for the interferometer shown in Fig.\,\ref{SGI} for different experimental parameters. The fixed parameters is $\nabla B=10^4$\,T/m in all the figures. The other parameters are set to be: $m_0=10^{-17}$\,kg (corresponds to $a_m=1.8\times10^{-2}\rm\,m/s^2$) and $t_e=1$\,s in the sub-figure (a), $m_0=10^{-17}$\,kg and $t_a=0.5$\,s in the sub-figure (b), and $t_a=0.5$\,s and $t_e=1$\,s in the sub-figure (c). As we have shown in all the figures, the transfer function $F(\omega)$ approaches a constant value in the low-frequency range, and decreases as a polynomial of $\omega$ in the high-frequency rregime. By comparing the sub-figures (a) with (b), we can find that the transfer function $F(\omega)$ is more sensitive to the value of the splitting time $t_a$, than the free-falling time $t_e$, especially in the low-frequency range. As we can see from the sub-figure (c), the transfer function $F(\omega)\propto m_0^{-4}$}. }
    \label{Fgg-symmetric}
\end{figure*}

We consider the specific configuration shown in Fig.\,\ref{SGI} ~\cite{Marshman:2018upe}. 
The interferometer is set to freely fall, and the creation and recombination stages control the superposition along the $x$-axis. For simplicity, the acceleration during the splitting and recombining parts is assumed to be constant, which can be achieved in a Stern-Gerlach apparatus with constant magnetic field gradient. The absolute value of the acceleration is given by, see~\cite{PhysRevLett.119.240401,Toros:2020dbf}.
\begin{equation}\label{am-magnetic}
    a_m = \frac{g\mu_B}{m_0}\vert\nabla B\vert,
\end{equation}
where $g=2$ is the Lande g-factor, $\mu_B=9\times10^{-24}$\,J/T is the Bohr magneton, $m_0$ is the mass of the interferometer and $\nabla B=10^4$\,T/m~\cite{Nat.Commun.4.2424, Marshman:2021wyk, Henkel:2021wmj} is the gradient of the magnetic field. The direction of the acceleration $a_m$ depends on the gradient of the magnetic field, and the value of the spin in each arm. The magnetic field gradient makes the system on the right path accelerate during $[0,t_a]$ and $[2t_a+t_e,3t_a+t_e]$, decelerate during $[t_a,2t_a]$ and $[3t_a+t_e,4t_a+t_e]$, while in the intermediate interval $[2t_a,2t_a+t_e]$ it is vanishingly small, while the part of the system on the left path is in free-fall. The transfer function for such an interferometer is given by \footnote{A similar form of the transfer function has been obtained also in~\cite{Toros:2020dbf} for  two symmetric interferometers located at distance $\pm d/2$ from the origin (i.e., a dual two matter-wave interferometers). Each interferometer is located asymmetrically with respect to the origin (i.e., either left or right of the origin). As the origin coincides with the center of the harmonic trap, each individual interferometer acquires different GGN induced phases on the two arms, leading to a GGN as a sensor in the combined dual two matter-wave interferometer. For more details, see Appendix~\ref{AppendixB}.}:
\begin{alignat}{2}
&  F(\omega) = &&16 \frac{a_m^4}{\omega^{10}}
\bigg(-t_a^2 \omega ^2 \sin \left(\omega  ( t_a+t_e/2)\right)\nonumber\\
& &&+\left(t_a^2 \omega ^2+3\right) \sin \left(t_e \omega /2\right)-3 \sin\left( \omega (2 t_a+t_e/2)\right)\nonumber\\
& &&+6 t_a \omega \cos \left(\omega(t_a+t_e/2)\right)\bigg)^2.
        \label{Fgg-half-eq}
\end{alignat}
The transfer function $F(\omega)$ is plotted in Fig.~\ref{Fgg-symmetric} with different values for the splitting time $t_a$, the free-falling time $t_e$, and the interferometer mass $m_0$.

As we have shown in sub-figures (a) and (b) of Fig.~\ref{Fgg-symmetric}, the splitting time, $t_a$, and the free-falling time, $t_e$, significantly affect on the behaviour of the transfer function $F(\omega)$. The splitting time has a greater impact on the absolute value of $F(\omega)$, while the free-falling time has a greater impact on the oscillatory behaviour of $F(\omega)$.

At low frequency, $\omega\ll2\pi/(4t_a+t_e)$, one can find that $F(\omega)$ reaches the constant value~\footnote{Using $\sin u \approx u-1/6 u^3$, and $\cos u \approx 1-1/2 u^2$, for $u\ll1$ in Eq.\,(\ref{Fgg-half-eq}), and introducing $\Delta x= a_m t_a^2$, which is the size of the superposition during the free-falling period.} $\Delta x^4 (23t_a+15t_e)^2/225$, which is much more sensitive to the value of $t_a$ than to the value of $t_e$. Setting $t_e=0$, we find a simple formula for the transfer function in the low frequency regime:
\begin{equation}
\bar{F}\equiv \lim_{\omega \rightarrow 0} F(\omega)= \frac{529}{225} \Delta x^4 t_a^2. \label{simple}
\end{equation}
In the high-frequency region, $\omega\gg2\pi/(4t_a+t_e)$, the transfer function $F(\omega)$ decreases rapidly as $\propto\omega^{-6}$. 

As we have shown in Fig.\,\ref{Fgg-symmetric} (c), the influence of the mass on the transfer function is a simple rescaling as $F(\omega) \propto m_0^{-4}$ according to Eqs.~\eqref{am-magnetic} and \eqref{Fgg-half-eq}. 
However, an interesting result is that for the configuration discussed in Appendix~\ref{AppendixB}, the corresponding transfer function $F(\omega)\propto m_0^{-2}$, which leads to  $\Gamma_{\rm noise}\propto m_0^2F(\omega)$, a mass-independent phase fluctuation.


\section{GGN in matter-wave interferometers}\label{section4}

In this section, we will analyse the phase fluctuation density due to the GGN.  In the Fermi normal coordinate system, constructed near the worldline of the laboratory~\cite{Poisson2011}, the Lagrangian in a non-relativistic limit is given by \cite{Toros:2020dbf}
\begin{equation}\label{free-falling Lagrangian}
    L_{\rm free-falling} = \frac{1}{2}m_0v^2 - m_0a_0x - \frac{1}{2}m_0 \underbrace{R_{0101}c^2}_{\equiv\omega_{gg}^2(t)}x^2,
\end{equation}
where the superposition direction is defined along the $x$-axis as shown in Fig.~\ref{SGI}. The first term on the right-hand side of Eq.~\eqref{free-falling Lagrangian} corresponds to a free-falling particle in a flat spacetime, and the other terms $m_0a_0x$ and $\frac{1}{2}m_0R_{0101}c^2x^2$ can be regarded as the acceleration noise and the GGN caused by the fluctuations in the metric, respectively~\cite{Toros:2020dbf}. 

For a free-falling experiment, the acceleration term $a_0$ will vanish according to the properties of the Fermi normal coordinates (in line with Einstein's equivalence principle), so this noise will be neglected in this paper. Therefore, we will solely focus on the noise $\omega_{gg}^2(t)$  in Eq.\,(\ref{free-falling Lagrangian}), which corresponds to the noise $\omega_{\rm noise}^2$ in Sec.~\ref{section2}.
As discussed, we characterize such a stochastic quantity by the noise PSD (see Eq.\,(\ref{Noise condition})). In particular, we introduce the GGN PSD, $S_{gg}(\omega)$, by the inverse-Fourier transformation, that is
\begin{alignat}{2}\label{Sgg-eq}
   & S_{gg}(\omega) = && \int\mathbb{E}[\omega_{gg}^2(t)\omega_{gg}^2(t+\tau)]\mathrm{e}^{i\omega\tau}d\tau\nonumber\\
  &  && \int\mathbb{E}[R_{0101}(t)R_{0101}(t+\tau)]c^4\mathrm{e}^{i\omega\tau}d\tau,
\end{alignat}
which has units of [$\rm Hz^4/Hz$]~\footnote{$S_{gg}(\omega)\sim\omega_{gg}^4/\omega$, where $\omega_{gg}$ describes the spacetime curvature noise and $\omega$ is the Fourier transformation frequency, so we write the unit as [$\rm Hz^4/Hz$] rather than [$\rm Hz^3$].}. 
There are many sources of GGN as noted in\cite{PhysRevD.58.122002, PhysRevD.60.082001, Beccaria:1998ap, Harms2019}, but in this paper we will focus on one particular source of GGN due to the smooth motion of external objects. In the next section we first adapt the two-dimensional classical analysis from ~\cite{PhysRevD.30.732} to matter-wave interferometry in three-spatial dimensions.


\section{Three dimensional GGN} \label{3D}

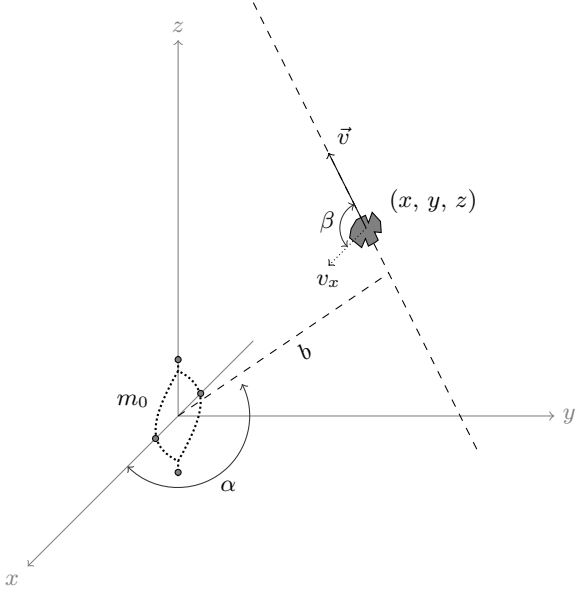
\begin{figure}
    \centering
    \begin{tikzpicture}
        \draw[->, gray] (0,0)--(5,0) node[right] {$y$};
        \draw[->, gray] (0,0)--(0,5) node[above] {$z$};
        \draw[->, gray] (1,1)--(-2,-2) node[below left] {$x$};
        \node at (-0.6, 0.2) {$m_0$};

        \draw[thick, densely dotted] (0,0.8)--(0,0.6);
        \draw[thick, densely dotted] (0,-0.8)--(0,-0.6);
        \draw[thick, densely dotted] (0,0.6) .. controls (-0.4,-0.1) and (-0.4,-0.4) .. (0,-0.6);
        \draw[thick, densely dotted] (0,-0.6) .. controls (0.4,0.1) and (0.4,0.4) .. (0,0.6);
        \draw[fill=gray] (0,0.75) circle (0.04);
        \draw[fill=gray] (0,-0.75) circle (0.04);
        \draw[fill=gray] (-0.3,-0.3) circle (0.04);
        \draw[fill=gray] (0.3,0.3) circle (0.04);

        \draw[dashed] (0,0)--node[below right, sloped] {$b$} (2.8,1.9);
        \draw[dashed] (1,5.5)--(4,-0.5);

        \coordinate (b) at (3,1.5);
        \coordinate (x) at (-2,-2);
        \coordinate (O) at (0,0);
        \pic["$\alpha$", draw=black!80, <->, angle eccentricity=1.2, angle radius=0.95cm] {angle=x--O--b};

        \draw[fill=gray] (2.69, 2.59) -- (2.58, 2.71) -- (2.53, 2.57) -- (2.49, 2.67) -- (2.37, 2.61) -- (2.3, 2.49) -- (2.28, 2.37) -- (2.44, 2.24) -- (2.49, 2.36) -- (2.53, 2.26) --  (2.62, 2.31) -- (2.66, 2.34) -- (2.61, 2.45) -- (2.7, 2.44) -- (2.69, 2.59) node[above right] {($x$, $y$, $z$)};
        \draw[->] (2.5,2.5)--(2,3.5) node[above right] {$\vec{v}$};

        \draw[densely dotted, ->] (2.5,2.5)--(2,2) node[below] {$v_x$};
        \coordinate (v) at (2,3.5);
        \coordinate (vx) at (2,2);
        \coordinate (P) at (2.5,2.5);
        \pic["$\beta$", draw=black!80, <->, angle eccentricity=1.5, angle radius=0.35cm] {angle=v--P--vx};
    \end{tikzpicture}
    \caption{\small {Three-dimensional GGN caused by the smooth motion of an external object. The external object is located at a point $(x,\ y,\ z)$ at time $t$, and moves with a constant velocity $\vec{v}=(v_x,v_y,v_z)$, while the interferometer of mass $m_0$ is located at the origin, with the superposition along the $x$-axis. The impact parameter is denoted here as $b$, and the projection angles are defined as $\cos\alpha=x_0/b$ and $\cos\beta=v_x/v$, where $x_0$ is the x-coordinate at $t=0$ and $v_x$ is the x-component of the constant velocity $\vec{v}$.}}
    \label{3-dim model}
\end{figure}
To quantify the achievable sensitivity for measuring the GGN in three spatial dimensions, we first compute the corresponding PSD $S_{gg}(\omega)$. Consider the model shown in Fig.~\ref{3-dim model}, and suppose that the external object whose coordinate is denoted by $\vec{r}=(x,\ y,\ z)$ moves with a uniform velocity $\vec{v}=(v_x,\ v_y,\ v_z)$, and with an impact parameter $b$. Then the local acceleration of the interferometer caused by the external mass at a given time, $t$, will be given by:
\begin{equation}
\begin{split}
    \vec{a}(t) &= \frac{GM}{r^2(t)}\frac{\vec{r}(t)}{r(t)} \\
        &= \frac{GM}{r^3(t)}x(t)\vec{e}_x + \frac{GM}{r^3(t)}y(t)\vec{e}_y + \frac{GM}{r^3(t)}z(t)\vec{e}_z,
\end{split}
\end{equation}
where $\vec{e}_j$ ($j=x,y,z$) are the unit basis vectors.
Since the external mass is assumed to be moving with a uniform velocity, one can write down $r^2(t)=b^2+v^2t^2$ and $x(t)=x_0+v_xt$ if $t=0$ is defined as the time when the external object is at the closest point. Further, if we introduce the projection angles 
\begin{equation}
\cos\alpha=x_0/b,~~\cos\beta=v_x/v, \label{coscos}
\end{equation}
then the $x$-direction component of the acceleration $\vec{a}$ can be written as
\begin{equation}
\begin{split}
    a_x(t) &= \frac{GM}{b^2}\frac{x_0/b+v_xt/b}{(1+v^2t^2/b^2)^{3/2}} \\
        &= \frac{GM}{b^2}\frac{\cos\alpha+(vt/b)\cos\beta}{(1+v^2t^2/b^2)^{3/2}}.
\end{split}
\end{equation}
Then in the frequency space, the Fourier transform of $a_x(t)$ is given by~\footnote{Note that the superposition of the interferometer is along the $x$-axis and hence we project the acceleration vector along this direction.}
\begin{equation}\label{a-omega-3dim}
\begin{split}
    a_x(\omega) = &\frac{GM}{b^2}\left(\frac{\omega b}{v}\right)\left[\frac{x_0}{v}K_1\left(\frac{\omega b}{v}\right)+i\frac{b}{v}\frac{v_x}{v}K_0\left(\frac{\omega b}{v}\right)\right] \\
    = &\frac{a_\text{loc} }{ \omega}u_\omega^2\left[\cos\alpha K_1\left(u_\omega\right)+i\cos\beta K_0\left(u_\omega\right)\right],
\end{split}
\end{equation}
where $K_0(\;\cdot\;)$ and $K_1(\;\cdot\;)$ are the modified Bessel functions. In the second line of Eq.~\eqref{a-omega-3dim} we have introduced the local acceleration, $a_{\rm loc}$, and the frequency-dependent dimensionless ratio, $u_\omega$, defined as
\begin{equation}
a_{\rm loc}\equiv GM/b^2, \;\;\; u_\omega\equiv\omega b/v, \label{definitions}
\end{equation}
which, as we will see, control the behaviour of the GGN.

\begin{figure}
    \includegraphics[scale=0.2]{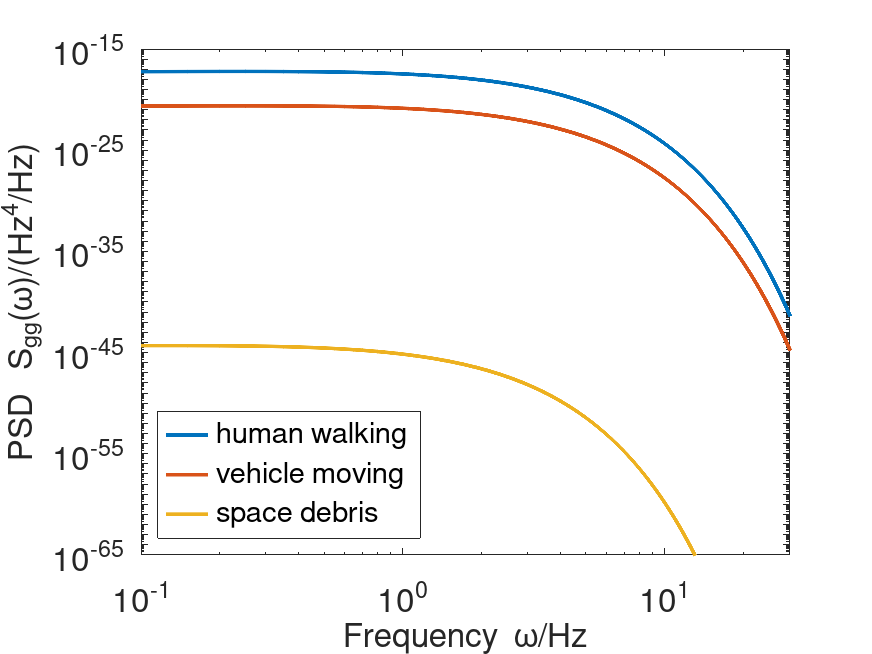}
    \caption{\small {We have shown the PSD of the GGN for several sources, including human walking, vehicles moving, and space debris according to Eq.\,(\ref{Sgg-3dim}). The masses are set as $50$\,kg, $10^3$\,kg and $10^3$\,kg in respect, the speeds are $1$\,m/s, $10$\,m/s and $5\times10^4$\,m/s respectively, and the impact parameter is set as $1$\,m, $10$\,m and $10^5$\,m in respect. As shown in the gravitational-wave literature~\cite{Harms2019, Acernese_2014, Bader_2022}, the GGN usually has a dominant contribution in the low-frequency range. The PSD for the GGN is usually smaller than $10^{-20}\,\rm Hz^4/Hz$~\cite{Acernese_2014, Bader_2022}, while it can reach $10^{-15}\,\rm Hz^4/Hz$ level for the human walking, and this is the reason why an ultra-sensitive experiment requires an exclusion zone for human activities~\cite{PhysRevD.30.732, PhysRevD.60.082001,Toros:2020dbf}. In this work, we, however, propose to detect such a tiny GGN as a signal by designing a suitable interferometer, i.e., by optimising the transfer function in Eq.~\eqref{Fgg-half-eq}. As we discuss in the text, by tuning the interferometric times, we can obtain a transfer function which can induce a detectable phase fluctuation, $\Gamma_{gg}$, in the specific frequency range.}}
    \label{Sgg-plot}
\end{figure}

The PSD of the acceleration noise on $a_x(\omega)$ can be computed as
~\footnote{According to the Wiener-Khinchin theorem, the PSD of $a_x(\omega)$ is given by $S_{aa}(\omega)=\int\mathbb{E}[a(t)a(t+\tau)]e^{i\omega\tau}d\tau$. The statistical average $\mathbb{E}[a(t)a(t+\tau)]$ can be calculated by time average $\mathbb{E}[a(t)a(t+\tau)]=\frac{1}{T}\int a(t)a(t+\tau)dt$. Then one can obtain the formula of PSD as 
\begin{equation*} 
\begin{split} S_{aa}(\omega)&=\frac{1}{T}\int\int a(t)a(t+\tau)e^{i\omega\tau}d\tau dt \\&=\frac{1}{T}\int\int a(t_1)a(t_2)e^{i\omega(t_1-t_2)}dt_1dt_2 \\&=\frac{1}{T}\left|\int a(t)e^{i\omega t}dt\right|^2=\frac{|a_x(\omega)|^2}{T}. 
\end{split}
\end{equation*}
}
\begin{equation} \label{Saa}
    S_{aa}(\omega) = \frac{|a_x(\omega)|^2}{T},
\end{equation}
where $T$ is the scattering time between the external mass and the interferometer (in this context, playing the role of the signal and sensor, respectively). A rough estimation of $T\sim b/v$, because the moving object is at a distance, ($r(t)=\sqrt{v^2t^2+b^2}$), that the interaction becomes negligible after $T\geq b/v$. An exact estimation of $T\sim b/v$ was also made in Ref.~\cite{PhysRevD.30.732}. We have particularly chosen the same estimation to match those results for two and three dimensions, discussed in the Appendix A. By combining Eqs.~\eqref{a-omega-3dim} and \eqref{Saa}, we can obtain the PSD for the acceleration noise,
\begin{equation}
    S_{aa}(\omega) = \frac{a_{\rm loc}^2}{\omega} u_\omega^3\big[ \cos^2\alpha K_1^2\left(u_\omega\right) +\cos^2\beta K_0^2\left(u_\omega\right)\big].
\end{equation}
Since the local acceleration $a_{\rm loc}$ is caused by the fluctuation of the local spacetime curvature, one may have the relation $a_{\rm loc} \sim R_{0101}c^2b$
\footnote{Consider the Newtonian potential, $V_G=\frac{GM_{ext}m}{b+\delta r}$, caused by an external mass $M_{\rm ext}$, where $\delta r$ is the fluctuation of the distance $b$. We can expand up to the second order, $V_G\sim G\frac{M_{\rm ext}m}{b} - G\frac{M_{\rm ext}m}{b^2}\delta r + G\frac{M_{\rm ext}m}{b^3}(\delta r)^2$. By comparing the Lagrangian of a freely-falling system, (\ref{free-falling Lagrangian}), we can obtain that $G\frac{M_{\rm ext}}{b^3}\sim \frac{1}{2}R_{0101}c^2$. Since, the local acceleration is caused by $M_{\rm ext}$, and $a_{\rm loc}=\frac{GM_{\rm ext}}{b^2}$, then we have $a_{\rm loc} \sim R_{0101}c^2b$.},
then the PSD for the local acceleration satisfies $S_{aa}(\omega)\sim S_{gg}(\omega)b^2$. Finally, the PSD of the GGN is given by
\begin{equation}\label{Sgg-3dim}
    S_{gg}(\omega) =  \frac{a_{\rm loc}^2}{\omega b^2} u_\omega^3 \left[\cos^2\alpha K_1^2\left(u_\omega\right)+\cos^2\beta K_0^2\left(u_\omega\right)\right].
\end{equation}
%


For example, the PSD of several sources such as human walking, vehicles moving, and space debris moving with a constant velocity is shown in Fig.\,\ref{Sgg-plot}. In gravitational-wave interferometers, $S_{gg}(\omega)$ is regarded as a source of noise, and is mitigated from $10^{-15}\,\rm Hz^4/Hz$ down to about $10^{-20}\,\rm Hz^4/Hz$ for human walking by setting a suitable exclusion zones\cite{Acernese_2014, Bader_2022, PhysRevD.30.732, PhysRevD.60.082001, Toros:2020dbf}.

We want to devise an interferometer that is capable of detecting weak GGN as signals in the low-frequency range by optimising the interferometric parameters. 
From Eqs.~\eqref{phase-fluctuation} and \eqref{Sgg-3dim}, we find that the the corresponding phase fluctuation is given by
\begin{alignat}{2}
    & \Gamma_{gg} = && \left(\frac{2m_0 a_{\rm loc}}{\hbar b}\right)^2  \  \int \frac{u_\omega^3 F(\omega)}{\omega} \nonumber\\
&    &&\left[\cos^2\alpha K_1^2\left(u_\omega\right)+\cos^2\beta K_0^2\left(u_\omega\right)\right]d\omega. \label{Gammagg}
\end{alignat}

Note that the PSD for the GGN $S_{gg}(\omega)$ approximately converges to zero in the low-frequency limit $\omega\to 0^+$, while the transfer function $F(\omega)$ converges to a non-zero constant, so the lower bound $\omega_{\rm min}=2\pi/t_{\rm exp}$ of the integration is not so relavant for the total phase fluctuation, $\Gamma_{gg}$. However, it still matters for some other sources of noise which diverges in the low frequency region, see ~\cite{Toros:2020dbf}. 

In experiments, the minimum measurable value of $\Gamma_{gg}$ will be determined by the the overall phase sensitivity. In the following we will assume $\Gamma_{gg}=0.01$ as a threshold value below which we can no longer reliably measure the phase fluctuations. Given such a threshold value for $\Gamma_{gg}$ we can then ask what should be the characteristic of the interferometer, such that it can discern a particular GGN as a signal. The interferometer mass, $m_0$, and the the superposition size, $\Delta x$, control the overall amplitude of the signal, while the beam-splitting time, $t_a$, and the free-fall time, $t_e$, control the sensitivity in a particular frequency range.
\begin{figure*}[ht!]
     \centering
     \begin{subfigure}[]{0.3\textwidth}
         \centering
         \includegraphics[width=\textwidth]{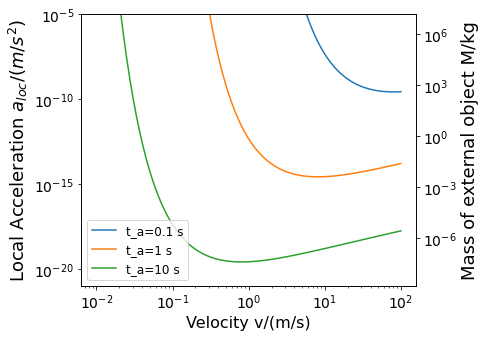}
         \subcaption*{(a)}
     \end{subfigure}
     \hfill
     \begin{subfigure}[]{0.3\textwidth}
         \centering
         \includegraphics[width=\textwidth]{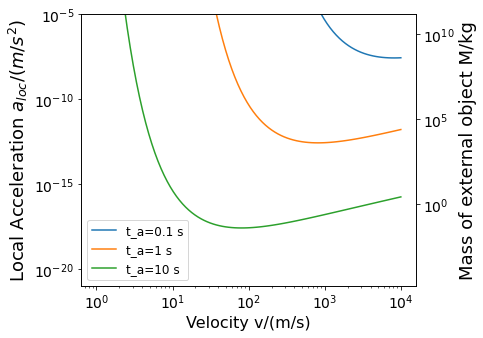}
         \subcaption*{(b)}
     \end{subfigure}
     \hfill
     \begin{subfigure}[]{0.3\textwidth}
         \centering
         \includegraphics[width=\textwidth]{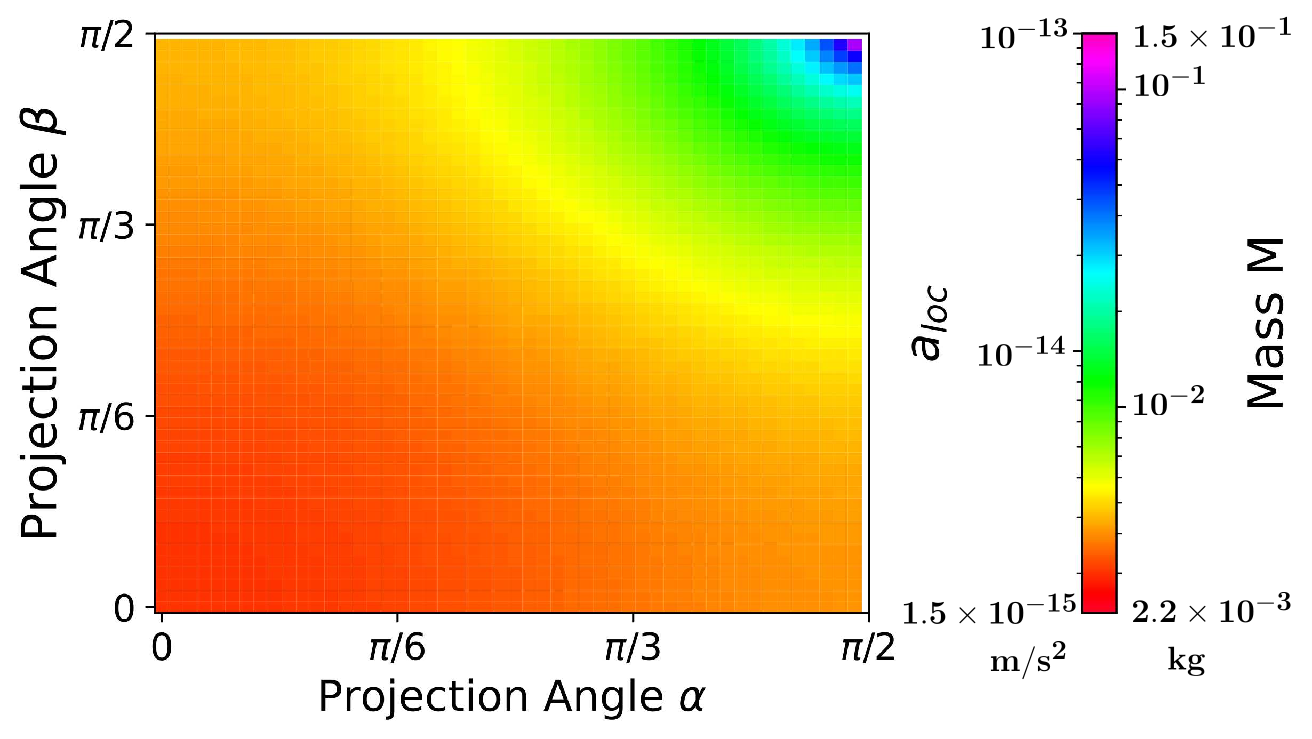}
         \subcaption*{(c)}
     \end{subfigure}
		\caption{\small {(a) The left-most panel shows the plot of the minimum detectable local acceleration, $a_\text{loc}=G M/b^2$ (left vertical-axis), or equivalently, the minimum detectable mass of the GGN source (right vertical-axis). The GGN source has an impact parameter $b=10$\,m and velocity v (horizontal-axis). The GGN sensitivity increases by increasing the beam-splitting time $t_a$ (the free-falling time $t_e=0$\,s for simplicity such that the total interferometric time is $4t_a$. The beam-splitting acceleration $a_m$ is set as $1.8\times10^{-2}\rm\,m/s^2$ (the corresponding magnetic field gradient we have used here is $\nabla B=10^4$\,T/m), and the mass of the interferometer is $m_0=10^{-17}$\,kg. (b) The middle panel shows similar to (a) but with the impact parameter set to $b=1000$\,m, requiring longer beam-splitting times $t_a$ to achieve the same sensitivity for the minimum detectable mass (corresponding to a better sensitivity of the minimum detectable local acceleration). (c) The right most panel shows the colour map for the minimum detectable local acceleration as a function of the projection angles $\alpha$ (horizontal-axis) and $\beta$ (vertical-axis) quantifying the relative orientation of the interferometer and the motion of the external object (see Fig.~\ref{3-dim model}). The impact factor has been set to $b=10$\,m and the velocity to $v=10$\,m/s. The optimal sensitivity is achieved when the motion of the external object is aligned with the axis of the interferometer ($\alpha=0$ and $\beta=0$).}}         
        \label{fig:lab}
\end{figure*}

From Eq.~\eqref{Gammagg} we can find the local gravitational acceleration
\begin{alignat}{2}\label{aloc-3dim}
   & a_{\rm loc}(M) =  \frac{\hbar b \sqrt{\Gamma_{gg}}}{2m_0}
   \bigg( &&\int \frac{u_\omega^3F(\omega)}{\omega}\big(\cos^2\alpha K_1^2\left(u_\omega\right) \nonumber\\
 &  && +\cos^2\beta K_0^2\left(u_\omega\right)\big)d\omega\bigg)^{-1/2},
\end{alignat}    
where the right-hand side fixes all the parameters, except the mass $M$ of the external object. Eq.~\eqref{aloc-3dim} thus provides a simple expression to estimate the minimum acceleration that one can sense given the threshold phase sensitivity $\Gamma_{gg}$. Since the impact parameter, $b$ is also fixed on the right-hand side of Eq.~\eqref{aloc-3dim} we find from Eq.~\eqref{definitions} that the minimum detectable mass $M$ of the external object with impact factor $b$ (moving with velocity $v$, and with its direction parametrised by the angles $\alpha$ and $\beta$), is given by $M=a_{\rm loc}b^2/G$.

\section{Sensing GGN sources in an earth-based laboratories and space-debris in the vicinity of satellites}\label{nearGGN} 

We now apply the model developed in the previous sections to sense GGN from two different types of sources. For simplicity we will set the free-fall time to $t_e=0$ and vary only the beam-splitting time $t_a$. We will focus on sensing GGN in the vicinity of Earth-based laboratories and sensing space-debris in the vicinity of satellites (Sec.~\ref{nearGGN}). The goal of this section is to check the feasibility of tracking the motion of the objects, ideally in real-time, and hence we consider the total experimental time to be the smallest possible, i.e., $t_\text{exp}=4t_a$. To make a statistically significant number of experimental runs we would thus need to consider an array of interferometers operating simultaneously.

Now we quantify the sensitivity to GGN signals caused by the motion of small objects in the proximity of experiments. As we will see, unknown light objects, even if moving at slow speeds, can be a significant source of GGN for state-of-the-art experiments, which become sensitive to tiny local accelerations. 

We first focus on GGN sources that could be present inside earth-based laboratories. In particular, we will consider external objects in the velocity range $(10^{-2}-10^{2})\,{\rm m/s}$, and with masses in the range from $(10^{-5} - 10^3)$\,kg. We will further assume that the external object, acting as the GGN source, has an impact factor $b=10$\,m. 

As discussed in Sec.~\ref{3D} we will set the GGN phase to the value $\Gamma_{gg} \geq 0.01$ \footnote{In a concrete experimental setup one has to estimate the achievable phase sensitivity by characterising various background noises. Here we have used the value $\Gamma_{gg} \geq 0.01$ is chosen as a concrete example (see comment below Eq.~\eqref{Gammagg}).}. If one fixes also the beam-splitting time $t_a$ one can then evaluate the local acceleration $a_{\rm loc}$. Using Eq.~\eqref{definitions} one can then readily determine also the minimum detectable mass $M$ of the GGN source. 

As shown in Fig.\,\ref{fig:lab} (a), when $v\to 0$ or $v \to \infty$, the local acceleration $a_{\rm loc}$ tends to infinity and the minimum detectable mass $M$ becomes extremely large. Indeed, when the external object moves too slowly or too fast, its GGN signal decreases as the frequency range of the interferometer $\sim t_a^{-1}$ is no longer compatible with the characteristic frequency of the GGN source given by $v/b$. The interferometer performs optimally as a GGN sensor when $t_a$ is comparable to $ b/v$.

A similar analysis as discussed above can be also adapted for sensing space debris in the vicinity of satellites~\cite{article, Cowardin2017CharacterizationOO}. For illustration, we will consider the debris at impact factor $b=1000$\,m and with velocity in the range $(10^{0}-10^4)\,{\rm m/s}$. We consider the same beam-splitting times as in the previous section, although the beam-splitting time could be significantly extended in space~\cite{aveline2020observation,gasbarri2021testing}. In Fig.\,\ref{fig:lab} (b) we show the measurable local acceleration, or equivalently, the minimum detectable mass of the GGN source. 

In Fig.\,\ref{fig:lab} (c) we also show the minimum detectable mass as a function of the projection, $\cos\alpha$ and $\cos\beta$, defined in Eq.~\eqref{coscos} evaluated for a fixed beam-splitting time $t_a$, fixed velocity $v=10$\,m/s, and fixed impact factor $b=10$\,m. The optimal sensitivity is achieved for $\cos\alpha=\cos\beta=1$ corresponding the external object moving along the $x$-axis.


\section{Summary}\label{section5}

In this paper, we first made a brief review of frequency-space analysis for matter-wave interferometry. We pointed out that the spectral density of the phase fluctuation caused by a noise can be always factorized into the noise part (described by the corresponding PSD) and the trajectory part (described by the so-called transfer function defined by Eq.\,(\ref{shape-function})). Although we have primarily focused on a SG scheme with nanoparticles, a similar analysis could be readily adapted to other types of matter-wave interferometers, such as those based on ultra-cold atom Bose-Einstein Condensate (BEC)\cite{Peters1999, PhysRevLett.120.183604, PhysRevLett.125.191101, gravitational_AB_effect, science.abm6854}.

We have developed a 3D model for the GGN signal of a moving external object, and obtained the corresponding PSD in Eq.\,(\ref{Sgg-3dim}),  generalizing the two dimensional model in \cite{PhysRevD.30.732}. Based on the PSD of gravity gradient signal, we then derived the expression Eq.\,(\ref{aloc-3dim}) and Eq.\,(\ref{aloc-3dim-approx}), which quantifies the local gravitational acceleration, or equivalently, the minimum detectable mass of the GGN source.

Finally, we applied the developed model to investigate two distinct GGN sources, namely, slow moving objects in Earth-based laboratories and space debris near satellites, and studied how the GGN signal varies with the velocity, distance, and orientation. 

Of course, there are numerous challenges to be met before we can realize experimentally such a quantum sensor. Creating large spatial superpositions and achieving the required coherence time with large masses is a formidable challenge. Nonetheless, we foresee that a nanoparticle matter-wave interferometer can have many novel technological applications, complementing the fundamental tests of Newton's law or detecting the quantum gravity induced entanglement.

\begin{acknowledgements}
We would like to thank Ryan Marshman for helpful discussions. 
M. Wu would like to thank the China Scholarship Council (CSC) for financial support. MT acknowledges funding by the Leverhulme Trust (RPG-2020-197). SB would like to acknowledge EPSRC grants No. EP/N031105/1 and EP/S000267/1. 
\end{acknowledgements}

\appendix

\section{Three Dimensional GGN and Reduction to two Dimensions}\label{appendix}

The model established in Sec.~\ref{3D} is a generalisation of a well-known model discussed in Ref.~\cite{PhysRevD.30.732}. In this appendix, we will discuss the special cases of the 3 dimensional model, and show how it reduces to the results of 1-dimensional model of Ref.~\cite{PhysRevD.30.732}.

\subsection{Three dimensional model}

When $\omega b/v\gg1$, we can make some approximations which are useful to investigate the slowly moving external objects (see Sec.~\ref{nearGGN}). In this latter regime, the modified Bessel functions can be approximated as
\begin{equation}\label{approx-Bessel}
    K_0\left(u_\omega\right)\approx K_1\left(u_\omega\right)\approx \sqrt{\frac{\pi}{2}}\frac{e^{-u_\omega}}{\sqrt{u_\omega}}.
\end{equation}
Then the PSD for the GGN in Eq.~\eqref{Sgg-3dim} can be reduced to:
\begin{equation}\label{Sgg-3dim-approx}
    S_{gg}(\omega) = \frac{a_\text{loc}^2}{\omega b^2}u_\omega^2(\cos^2\alpha+\cos^2\beta)e^{-2 u_\omega}.
\end{equation}
Note that when, $\alpha=0$, and, $\beta=\pi/2$, the PSD can be further reduced to
\begin{equation}
    S_{gg}(\omega) = \frac{a_\text{loc}^2}{\omega b^2}u_\omega^2e^{-2 u_\omega},
\end{equation}
which is the same result in Ref.~\cite{PhysRevD.30.732}. 
Based on the reduced PSD in Eq.~\eqref{Sgg-3dim-approx}, the local acceleration, Eq.\,(\ref{aloc-3dim}), can be simplified to
\begin{equation}\label{aloc-3dim-approx}
    a_{\rm loc} = \frac{\hbar v}{2m_0}\sqrt{\frac{\Gamma_{gg}}{(\cos^2\alpha+\cos^2\beta)\int \omega F(\omega)e^{-2 u_\omega} d\omega}}.
\end{equation}
Physically, the condition, $\omega b/v\geq1$, gives, $b/v\geq1/\omega_{\rm min}\sim t_{\rm exp}$,
which constrains the interaction time, $T\sim b/v$, to be longer than the interferometric times, $t_{a},~t_{e}$. For example, a walking person who is moving with the speed $\sim1$\,m/s, at a distance, $\sim1$\,m, so the corresponding ratio $b/v\sim1$\,s satisfies the condition $b/v\geq t_{\rm e},t_{\rm a}\sim1$\,s. 
 
However, the approximation in Eq.~\eqref{aloc-3dim-approx} gives reasonable values as long as we are in the regime $\tilde{\omega}_j b/v\gg1$, where $\tilde{\omega}_j=2\pi/t_j$ ($j=a,e$) denotes the characteristic frequencies of the interferometer. The latter regime has the following hierarchy of times:
\begin{equation}
  t_a,t_e \ll b/v  \ll t_\text{exp}, \label{hierar}
\end{equation} 
where we recall that $t_\text{exp}$ is the total experimental time, 
$b/v$ can be interpreted as the interaction time, and $ t_a,t_e$ are the beam-splitting time and free-evolution time of a single interferometric loop, respectively. In such a regime we can make the approximation $F(\omega)\approx \bar{F}$, where $\bar{F}$ is defined in Eq.~\eqref{simple}. The integrations in Eq.~\eqref{aloc-3dim} then reduce to
\begin{alignat}{2}
&\int_0^\infty u_\omega^2 K_0^2\left(u_\omega\right) d u_\omega &&= \frac{\pi^2}{32}\approx 0.31, \label{00}\\
&\int_0^\infty u_\omega^2 K_1^2\left(u_\omega\right) d u_\omega &&= \frac{3\pi^2}{32}\approx 0.93, \label{11}
\end{alignat}
where we have changed the integration variable to $u_\omega=b\omega/v$ defined in Eq.~\eqref{definitions}. On the other hand, using the approximation in Eq.~\eqref{approx-Bessel}, the relevant integration in Eq.~\eqref{aloc-3dim} evaluates to:
\begin{equation}
\int_0^\infty u_\omega \left(\sqrt{\frac{\pi}{2}}\frac{e^{-u_\omega}}{\sqrt{u_\omega}}\right)^2 d u_\omega= \frac{\pi}{8}\approx 0.40.
\end{equation}
which is of the same order of magnitude as the results obtained in Eqs.~\eqref{00} and \eqref{11}. Since in this work we are primarily interested in the order of magnitude estimates, we will thus use the approximation in Eq.~\eqref{aloc-3dim-approx} also for the regime given in Eq.~\eqref{hierar}.

\subsection{GGN in two dimensions}\label{appendixA}

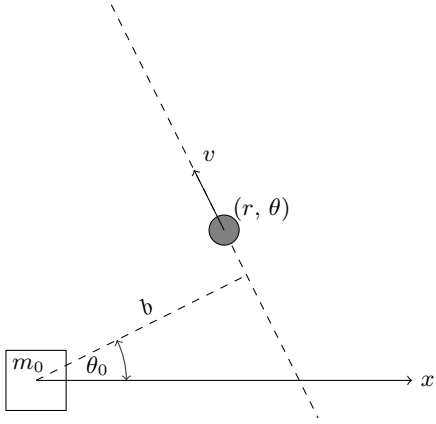
\begin{figure}
    \begin{tikzpicture}
        \draw[->] (0,0)--(5,0) node[right] {$x$};
        \draw[dashed] (0,0)--node[above right, sloped] {$b$} (2.8,1.4);
        \draw (-0.4,-0.4) rectangle (0.4,0.4);
        \node at (-0.1, 0.2) {$m_0$};
    
        \coordinate (A1) at (2,0);
        \coordinate (A2) at (2,1);
        \coordinate (O) at (0,0);
        \pic["$\theta_0$", draw=black!80, <->, angle eccentricity=0.7, angle radius=1.2cm] {angle=A1--O--A2};
    
        \draw[dashed] (1,5)--(3.75,-0.5);
        \draw[fill=gray] (2.5,2) circle (0.2) node[above right] {($r$, $\theta$)};
        \draw[->] (2.5,2)--(2.1,2.8) node[above right] {$v$};
    \end{tikzpicture}
    \caption{\small The two dimensional  model for the GGN caused by a smooth motion. The external object is originally located at point $(b,\ \theta_0)$, and moves with a constant speed $v$. }
    \label{2-dim model}
    \end{figure}

Now we will show how the three dimensional model developed in Sec.~\ref{3D} reduces to a two-dimensional model when the external object and the quantum sensor are confined to a plane (see Fig.\,\ref{2-dim model}). Comparing to the three dimensional model from the main text, we only need one polar angle $\theta$ to describe the motion of the external object moving at impact factor $b$. As we will see below, if we further set the angle to $\theta_0=0$, then the two dimensional model reduces to the original model proposed in \cite{PhysRevD.30.732}. The acceleration caused by the Newtonian force in the x-direction is given by:
\begin{equation}
    a_x(t) = \frac{GM}{b^2}\frac{1}{(1+(vt/b)^2)^{3/2}}(\cos\theta_0+(vt/b)\sin\theta_0),
\end{equation}
so in the frequency space, the local acceleration is
\begin{equation}
    a_x(\omega) = \frac{GM}{b^2\omega} u_\omega^2\left(\cos\theta_0 K_1\left(u_\omega\right)+i \sin\theta_0 K_0\left(u_\omega\right)\right),
\end{equation}
where $K_\alpha(\;\cdot\;)$ is the modified Bessel function, and we have introduced $u_\omega=b\omega/v$ (see Eq.~\eqref{definitions} in the main text). Comparing to the three dimensional result in Eq.\,(\ref{a-omega-3dim}), the projection angle $\alpha$ and $\beta$ becomes $\theta_0$ and $\pi/2-\theta_0$, respectively. 

According to $S_{aa}(\omega)=|a_x(\omega)|^2/T$, $T=b/v$, and $S_{gg}(\omega)=S_{aa}(\omega)/b^2$, the PSD for the GGN in the two dimensional case is given by:
\begin{equation}
    S_{gg}(\omega) = \frac{a_\text{loc}^2 u_\omega^3}{\omega b^2 }  \left[\cos^2\theta_0 K_1^2\left(u_\omega\right)+\sin^2\theta_0 K_0^2\left(u_\omega\right)\right], \label{sgg2}
\end{equation}  
where we have introduced, $a_\text{loc}=G M/b^2$ (see Eq.~\eqref{definitions} in the main text). The corresponding phase fluctuation is given by:
\begin{alignat}{2}
    & \Gamma_{gg} = && \left(\frac{2m_0 a_{\rm loc}}{\hbar b}\right)^2  \  \int \frac{u_\omega^3 F(\omega)}{\omega} \nonumber\\
&    &&\left[\cos^2\theta_0 K_1^2\left(u_\omega \right)+\sin^2\theta_0 K_0^2\left( u_\omega \right)\right]d\omega. \label{Gammagg2}
\end{alignat}
From Eq.~\eqref{Gammagg2} we then readily find the local acceleration:
\begin{alignat}{2}\label{aloc-3dim22}
   & a_{\rm loc}(M) =  \frac{\hbar b \sqrt{\Gamma_{gg}}}{2m_0}
   \bigg( &&\int \frac{u_\omega^3F(\omega)}{\omega}\big(\cos^2\theta_0 K_1^2\left(u_\omega\right) \nonumber\\
 &  && +\sin^2\theta_0 K_0^2\left(u_\omega\right)\big)d\omega\bigg)^{-1/2}.
\end{alignat}    

If we now set $\theta_0=0$, we recover the result presented in \cite{PhysRevD.30.732}. In the regime, $u_\omega\gg1$, the modified Bessel's function can be approximated as $K_0(u_\omega)\sim K_1(u_\omega)\sim e^{-u_\omega}/u_\omega^{1/2}$ (see Eq.~\eqref{approx-Bessel}). In this regime, the PSD for the GGN in Eq.~\eqref{sgg2} reduces to
\begin{equation}
    S_{gg}(\omega) = \frac{a_\text{loc}^2}{b^2\omega} u_\omega^2e^{-2\omega b/v}. \label{112}
\end{equation}%
The GGN formula Eq.~\eqref{112} remains a decent approximation even when $u_\omega\sim1$ which is the regime considered in ~\cite{PhysRevD.30.732} where they have omitted the dimensionless prefactor $u_\omega^2$. Besides, as is seen in \eqref{112}, the choice of $T$ should be $b/v$ to match the result in \cite{PhysRevD.30.732}, otherwise there will be an additional factor. Finally, using Eq.~\eqref{112} we find that the local acceleration simplifies to the simple expression:
\begin{equation}
    a_{\rm loc} = \frac{\hbar v}{2m_0}\sqrt{\frac{\Gamma_{gg}}{\int \omega F(\omega)e^{-2 u_\omega} d\omega}}, \label{222}
\end{equation}
which matches Eq.~\eqref{aloc-3dim-approx} for $\alpha=0$ and $\beta=\pi/2$.

\section{GGN with two symmetric interferometers}\label{AppendixB}

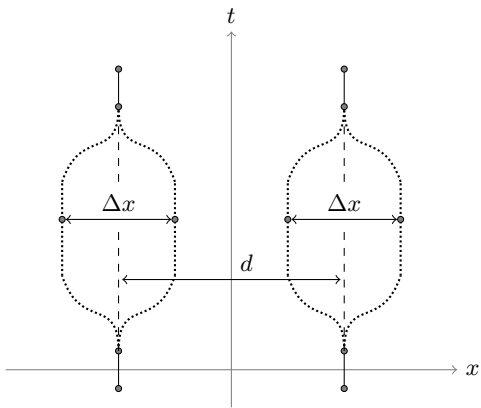
\begin{figure}
    \captionsetup{justification=justified}
    \begin{tikzpicture}
       \draw[->, gray] (-3,0)--(3,0) node[right, black] {$x$};
       \draw[->, gray] (0,-0.5)--(0,4.5) node[above, black] {$t$};
   
       \draw (1.5,-0.25)--(1.5,0.25);
       \draw[thick, densely dotted] (1.5,0.25) .. controls (1.5,1) and (1,0.5) .. (0.75,1.25);
       \draw[thick, densely dotted] (1.5,0.25) .. controls (1.5,1) and (2,0.5) .. (2.25,1.25);
       \draw[thick, densely dotted] (0.75,1.25)--(0.75,2.5);
       \draw[thick, densely dotted] (2.25,1.25)--(2.25,2.5);
       \draw[thick, densely dotted] (0.75,2.5) .. controls (1,3.25) and (1.5,2.75) .. (1.5,3.5);
       \draw[thick, densely dotted] (2.25,2.5) .. controls (2,3.25) and (1.5,2.75) .. (1.5,3.5);
       \draw (1.5,3.5)--(1.5,4);
       
       \draw (-1.5,-0.25)--(-1.5,0.25);
       \draw[thick, densely dotted] (-1.5,0.25) .. controls (-1.5,1) and (-1,0.5) .. (-0.75,1.25);
       \draw[thick, densely dotted] (-1.5,0.25) .. controls (-1.5,1) and (-2,0.5) .. (-2.25,1.25);
       \draw[thick, densely dotted] (-0.75,1.25)--(-0.75,2.5);
       \draw[thick, densely dotted] (-2.25,1.25)--(-2.25,2.5);
       \draw[thick, densely dotted] (-0.75,2.5) .. controls (-1,3.25) and (-1.5,2.75) .. (-1.5,3.5);
       \draw[thick, densely dotted] (-2.25,2.5) .. controls (-2,3.25) and (-1.5,2.75) .. (-1.5,3.5);
       \draw (-1.5,3.5)--(-1.5,4);

       \draw[fill=gray] (1.5,-0.25) circle (0.04);
       \draw[fill=gray] (1.5,0.25) circle (0.04);
       \draw[fill=gray] (0.75,2) circle (0.04);
       \draw[fill=gray] (2.25,2) circle (0.04);
       \draw[fill=gray] (1.5,3.5) circle (0.04);
       \draw[fill=gray] (1.5,4) circle (0.04);
       \draw[<->] (0.8,2)-- node[above]{$\Delta x$} (2.2,2);
   
       \draw[fill=gray] (-1.5,-0.25) circle (0.04);
       \draw[fill=gray] (-1.5,0.25) circle (0.04);
       \draw[fill=gray] (-0.75,2) circle (0.04);
       \draw[fill=gray] (-2.25,2) circle (0.04);
       \draw[fill=gray] (-1.5,3.5) circle (0.04);
       \draw[fill=gray] (-1.5,4) circle (0.04);
       \draw[<->] (-0.8,2)-- node[above]{$\Delta x$} (-2.2,2);
   
       \draw[dashed] (1.5,0.25)--(1.5,1.85);
       \draw[dashed] (1.5,2.5)--(1.5,3.25);
       \draw[dashed] (-1.5,0.25)--(-1.5,1.85);
       \draw[dashed] (-1.5,2.5)--(-1.5,3.25);
       \draw[<->] (-1.45,1.2)-- node[above right]{$d$} (1.45,1.2);
   \end{tikzpicture}
       \caption{\small {Illustration of the paths of the dual interferometer. A single symmetric interferometer is by itself not susceptible to sense GGN, because one can always choose the origin of the coordinate system, corresponding to the origin of the harmonic GGN metric perturbation, such that the paths are located symmetrically on each side (for example, the paths of the right interferometer are symmetric with respect to $x=d/2$ and the phases on each arm would become proportional to $\propto (\pm\Delta x/2 )^2$ -- one only generates an undetectable global phase on an individual interferometer). However, if one is considering joint observables of multiple interferometers placed along the $x$-axis, one can no longer make the phases on an individual interferometer $\propto x^2$ equal, leading to a nonzero GGN signal. Indeed, in the picture the phases on the two arms of the left (right) interferometer are given by $\propto (-d/2\pm\Delta x/2 )^2$ ($\propto (d/2\pm\Delta x/2 )^2$). More generally, two or more adjacent symmetric interferometers can become sensitive to GGN when the two arms of an individual individual interferometer are placed asymmetrically with respect to the origin of the local GGN metric pertubation \cite{Toros:2020dbf}. $d$ is the distance between the two interferometers, and $\Delta x$ is the superposition size, assumed equal for both interferometers.}}
       \label{SGI2}
   \end{figure}

   \begin{figure*}[ht!]
        \centering
        \begin{subfigure}[]{0.3\textwidth}
            \centering
            \includegraphics[width=\textwidth]{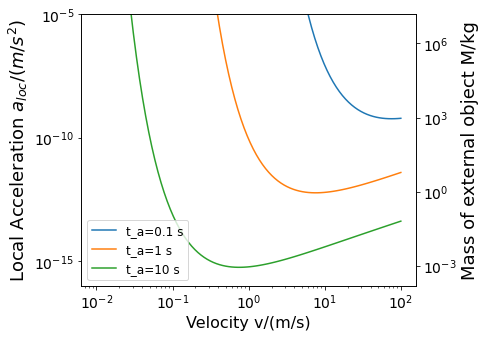}
            \subcaption*{(a)}
        \end{subfigure}
        \hfill
        \begin{subfigure}[]{0.3\textwidth}
            \centering
            \includegraphics[width=\textwidth]{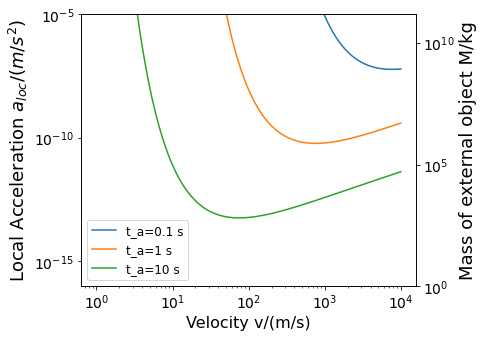}
            \subcaption*{(b)}
        \end{subfigure}
        \hfill
        \begin{subfigure}[]{0.3\textwidth}
            \centering
            \includegraphics[width=\textwidth]{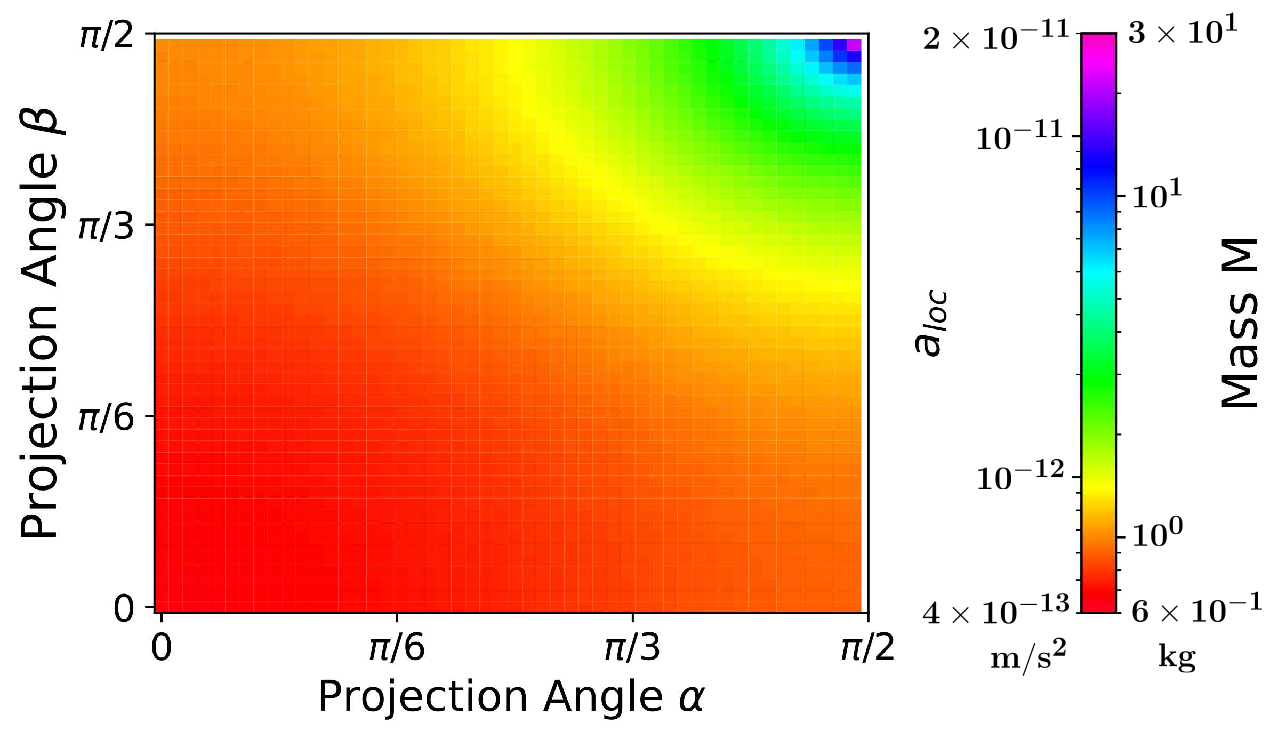}
            \subcaption*{(c)}
        \end{subfigure}
           \caption{\small {Same as Fig.~\ref{fig:lab} but for the dual interferometer in Fig.~\ref{SGI2} with the transfer function given in Eq.~\eqref{Fgg-syymetric-eq}. The mass of an individual interferometer is $m_0=10^{-17}$\,kg, and we have set the distance between the two interferometers to $d=100\,\mu$m, the splitting time $t_a=1$\,s, the free-falling time $t_e=$0\,s and the beam-splitting acceleration $a_m=1.8\times10^{-2}\rm\,m/s^2$ corresponding to the magnetic field gradient $\nabla B=10^4$\,T/m. (a) Plot of minimum detectable local acceleration, or equivalently, of the minimum detectable mass of the GGN source. The impact factor is set to $b=10$\,m. (b) Same as (a) but with the impact parameter set to $b=1000$\,m. (c) Colour map for the minimum detectable local acceleration as a function of the projection angles $\alpha$ and $\beta$. The impact factor of the external object has been set to $b=10$\,m and the velocity to $v=10$\,m/s.}}         
           \label{fig:lab2}
   \end{figure*}

For completeness we discuss the dual QGEM interferometer depicted in Fig.~\ref{SGI2}. Each individual interferometer (the left one or the right one) has the paths located asymmetrically with respect to the origin -- as such, the two paths of an individual interferometer acquire a nonzero phase difference from the harmonic trap generated by a GGN signal centered at the origin. In case, one is looking at joint properties of the two interferometers, such as an entanglement witness, the dual interferometer becomes sensitive to GGN~\cite{Toros:2020dbf}. 

The transfer function for symmetric interferometer is given by~\cite{Toros:2020dbf}:
\begin{equation}\label{Fgg-syymetric-eq}
    F(\omega) = 64d^2a_m^2\frac{\sin^4(\frac{\omega t_a}{2})\sin^2(\frac{1}{2}\omega(2t_a+t_e))}{\omega^6}.
\end{equation}
where $d$ denotes the distance between the centers of two interferometers (the rest of the parameters have the same meaning to the ones defined in the main text). 

An interesting observation is that the transfer function for this configuration is proportional to $m_0^{-2}$ rather than $m_0^{-4}$ in Eq.\,(\ref{Fgg-half-eq}). As a consequence the corresponding phase fluctuation density $\Gamma_{\rm noise}$ will be independent of $m_0$, according to Eq.\,(\ref{phase-fluctuation}). Thus, the mass of the superposition can be chosen arbitrarily for this configuration, which is an advantage.
We have discussed the minimum local acceleration, or equivalently, the minimum detectable mass, from sensing GGN in Fig.~\ref{fig:lab2}. We note that the dual QGEM interferometer is less sensitive to sense the GGN in comparison to the asymmetric MIMAC interferometer.

\bibliographystyle{apsrev}

\end{document}